\documentclass[prl,aps,showpacs]{revtex4}
\begin{document}
\draft
\title{Change of $^7$Be decay rate in exohedral and endohedral C$_{60}$ fullerene compounds and its implications}
\author{A. Ray$^{a}$, P. Das$^{a}$, S. K. Saha$^{b}$, S. K. Das$^{b}$,
J. J. Das$^{c}$, N. Madhavan$^{c}$, S. Nath$^{c}$, P. Sugathan$^{c}$,  
P. V. M. Rao$^{c,d}$, A. Jhingan$^{c}$}
\address{$^{a}$Variable Energy Cyclotron Centre, 1/AF, Bidhannagar,
Kolkata- 700064, India}
\address{$^{b}$Radiochemistry Division, Variable Energy Cyclotron Centre,
1/AF, Bidhannagar, Kolkata- 700064, India}
\address{$^{c}$Nuclear Science Centre, New Delhi, India}
\address{$^{d}$Department of Nuclear Physics, Andhra University,
Visakhapatnam - 530003, India}
\date{ September 15, 2005}

\begin{abstract}
The half-life of
$^{7}$Be implanted in a C$_{60}$ fullerene pellet and gold foil has been measured to be about the
same within $\approx$0.2$\%$. Using a radiochemical technique, we also measured that the probability of formation of endohedral $^{7}$Be@C$_{60}$ by nuclear implantation technique was (5.6$\pm$0.45)$\%$. It is known from earlier works that the half-life of endohedral $^{7}$Be@C$_{60}$ is $\approx$1.2$\%$ shorter than that of $^{7}$Be implanted in gold.  An analysis of these results
using linear muffin-tin orbital method calculations indicates that most of the implanted $^{7}$Be ions in fullerene C$_{60}$ stay at a distance of $\approx$5.3 {\AA} from the centers of nearest C$_{60}$ molecules forming exohedral compounds and those who enter the fullerene cages go to the centers of the cages forming 
endohedral $^{7}$Be@C$_{60}$ compounds.

\end{abstract}
\pacs{21.10.Tg,23.40.Hc,61.48.+c,23.40.-s,36.40.-c}
\maketitle

It is known from earlier works\cite{R1,R2,R3,R4, R5, R6} that the rate of orbital electron capture of $^{7}$Be is susceptible to the surrounding environment and depends on both the lattice structure and electron affinity of the host atoms. Calculations\cite{R2,R5, R6} also showed that the decay rate of $^{7}$Be should depend on its position in the host lattice. So the decay rate of $^{7}$Be in an atomic cluster such as fullerene C$_{60}$ or large biomolecule should also depend on its position with respect to the molecule and could be used as a tool to learn about the position of the implanted radioactive ion in the atomic cluster. In the future, this kind of study might also tell us about any abnormal change of DNA molecule in a living cell. 
   
It is already well-known\cite{R7,R8, R9} 
that different types of metal atoms(Be, Kr, Xe etc.) can be
inserted into C$_{60}$ fullerene cage forming endohedral compound 
by nuclear implantation technique. 
Many theoretical studies\cite{R10, R11, R12, R13, R14} have been done regarding the charge and equilibrium positions
of the implanted ion in both endohedral and exohedral fullerene complexes, but 
there is no corresponding experimental measurement. So it is 
important to address these questions experimentally. By comparing the measured half-lives of exohedral $^{7}$Be-C$_{60}$ and endohedral $^{7}$Be@C$_{60}$ with that of implanted $^{7}$Be in another well understood material such as gold, we can learn about the charge and equilibrium positions of $^{7}$Be in endohedral and exohedral fullerene C$_{60}$ complexes. 

In this work, we have measured the difference of half-lives of 
$^{7}$Be implanted in a gold (Au) foil and fullerene (C$_{60}$) pellet. 
A 25$\mu$m thick gold(Au) foil and a
500$\mu$m thick fullerene(C$_{60}$) pellet were bombarded by a 18 MeV
$^{7}$Be beam from Nuclear Science Center, New Delhi, India. The ranges
of 18 MeV $^{7}$Be in Au and C$_{60}$ are 12 $\mu$m and 
40$\mu$m respectively. The intensity of the $^{7}$Be beam was 
about 40,000 particles/sec and the duration of the irradiation 24 hours
for each sample. In order to obtain a 18 MeV $^{7}$Be beam, a liquid 
nitrogen cooled hydrogen gas cell at one atmospheric pressure 
was bombarded with a 15 pnA
21 MeV $^{7}$Li beam obtained from the pelletron machine of Nuclear Science
Center, New Delhi, India. The $^{7}$Be nuclei produced by the reaction
$^1H(^7Li,^7Be)^1n$ at 0$^o$ were separated from the
primary $^{7}$Li beam by using a recoil mass spectrometer called 
Heavy Ion Reaction Analyzer (HIRA) \cite{R15} operated in a suitable
ion optical mode\cite{R16}. Primary $^{7}$Li beam particles were rejected
by a slit system installed at the intermediate focal plane of the
spectrometer. The primary beam rejection factor was about 10$^{12}$ and the purity of the separated $^{7}$Be beam was about 92$\%$. The advantages of rejecting the primary $^{7}$Li beam and using only a high purity $^{7}$Be beam for implantation are to minimize the radiation damage of the sample and avoid production of unwanted radioactivity.

The $^{7}$Be implanted samples were brought to 
Variable Energy Cyclotron Center, Kolkata, India for off-line counting. 
Following electron capture, a $^{7}$Be nucleus has a 10.4$\%$ probability\cite{R17}
of populating the first excited state of $^{7}$Li which decays subsequently 
to its ground state emitting a 478-keV $\gamma$-ray photon.
The half-life of $^7$Be was determined 
by monitoring the intensity of this 478 keV $\gamma$-ray line with time. 
Two HPGe detectors (detector-1 and detector-2) having efficiency 
$\approx$30$\%$ were used to count the samples ($^{7}$Be
implanted Au foil and $^{7}$Be implanted fullerene pellet). In addition, 
a standard $^{133}$Ba source was also placed in front of each 
HPGe detector. The detectors were well shielded by lead bricks to
avoid any cross-talk between them and also to reduce the background
level. 
A typical $\gamma$-ray spectrum (for $^{7}$Be in C$_{60}$ fullerene) as 
recorded by detector-1 is shown in Fig.\ \ref{f1}. 
Apart from 478 keV $\gamma$-ray line coming from the decay of $^7$Be and other $\gamma$-ray lines from $^{133}$Ba, we also see  standard background $\gamma$-rays such as 
239 keV, 511 keV, 583 keV, 609 keV, 727 keV, 911 keV lines,  
but there was no other contaminant. The $\gamma$-ray spectrum of
$^{7}$Be in Au is very similar to Fig. 1. 

The count rate of 478 keV $\gamma$-ray was about 1 count per sec at the 
beginning of the run. Both the HPGe detectors were started at the same time, 
data was accumulated for 24 hours, stored in a computer 
and the spectra were cleared and the counting restarted. After counting for 7 days,
the positions of the samples were interchanged and counted again. This was done
to take care of any systematic error. The counting was continued for about 6
months. 

From each day's spectra, we determined the counts under 478 keV 
(N$_{\gamma}$(478)) and 356 keV (N$_{\gamma}$(356))photo-peaks
coming from $^7$Be and  $^{133}$Ba respectively.
The ratio N$_{\gamma}$(478)/N$_{\gamma}$(356) should be independent of
computer dead time and systematic errors and decay exponentially with time. In the case of data sets for $^{7}$Be in fullerene C$_{60}$ and $^{7}$Be in Au samples taken with detector-1, the reduced chisquare values of the exponential fits are 1.5 and 1.6 respectively. 
We also have similar data taken by detector-2
for $^{7}$Be in fullerene C$_{60}$ and Au and the reduced chisquare values of these exponential fits are 2.7 and 2.9 respectively. The $\gamma$-ray line shape of detector-2 showed slight tailings on both sides and this is probably responsible for obtaining comparatively poor exponential fit from the data set recorded by detector-2. So we compared
data sets taken by the same detector and analysed the same way with similar
quality of fits (reduced chisquare values) to cancel out the effect of any systematic error. We obtain the percentage
difference of the two decay rates i.e
$\frac{\lambda(Au)-\lambda(C_{60})}{\lambda(Au)}$=$(0.060\pm0.405)\%$ and
$(0.087\pm0.264)\%$ for detector-1 and detector-2 respectively, where only statistical errors have been considered.
Taking weighted average,
we finally obtain the percentage difference in decay rates of $^{7}$Be in 
gold and fullerene to be
$\frac{\lambda(Au)-\lambda(C_{60})}{\lambda(Au)}$=$(0.079\pm0.221)\%$.
So we find that the 
half-life of $^{7}$Be in gold and fullerene is about the same within
$\approx$0.2$\%$. 

In order to determine the half-life of $^{7}$Be in Au, we used known half-life\cite{R17} of $^{133}$Ba =(3836$\pm$15)days and only the data set taken by detector-1. The absolute half-life of $^{7}$Be in Au as obtained from the data set (reduced chisquare value = 1.6 for exponential fit) taken by detector-1 is = (53.60$\pm$0.19)days, where only statistical errors have been considered. The half-life of $^{7}$Be in Au as obtained from the data set taken by detector-2 agrees very well with that obtained from detector-1. However we think it would not be appropriate to reduce error bar on the absolute value of half-life by combining results from two data sets which give significantly different reduced chisquare values when fitted with exponential functions. Norman et al.\cite{R3}obtained half-life of $^{7}$Be in Au = (53.311$\pm$0.041)days and the reduced chisquare value for exponential fit of their data set was = 1.04. However in their experiment, a primary $^{7}$Li beam was incident on a Au foil and so some radiation damage of the Au lattice was expected. We also performed a similar experiment by bombarding a kapton foil with a 48 MeV $^{7}$Li beam from BARC-TIFR pelletron machine and implanting recoiled $^{7}$Be ions emitted at 0$^{\circ}$ in a Au foil placed behind the target. The implanted sample along with a $^{133}$Ba source were counted using a HPGe detector for about two months and the half-life of $^{7}$Be was determined by monitoring the ratio of counts under 478 keV and 356 keV $\gamma$-ray photo-peaks with time. The reduced chisquare of the exponential fit was = 1.45 and taking the half-life\cite{R17} of $^{133}$Ba =(3836$\pm$15)days, we obtained the half-life of $^{7}$Be in Au = (53.328$\pm$0.082)days, in good agreement with Norman et al.'s \cite{R3}result. So we think the radiation damage of Au lattice caused by a high flux incident primary heavy ion beam might be responsible\cite{R18}for slightly lowering (0.54$\pm$0.36)$\%$ the half-life of $^{7}$Be in Au. 

In a subsequent experiment, we performed a radiochemical separation of endohedral $^{7}$Be@C$_{60}$ compound. $^{7}$Be implanted fullerene C$_{60}$ catcher was dissolved in 5 ml 1,2,4 trichlorobenzene. Equal volume of 6N hydrochloric acid containing carriers was added to the solution and the mixture was thoroughly shaken for 5 minutes in a separating funnel. The organic and the aqueous phases were separated and the activities of both the phases were determined accurately by $\gamma$-spectroscopy using a 20$\%$ HPGe detector. The aqueous fraction was again mixed with equal volume of organic solvent and a second extraction was carried out. The second organic fraction contained $\le$4$\%$ of radioactivity compared to the first extraction. The organic fraction was also filtered through a millipore filter (pore size = 0.45 mm) to remove any insoluble material. No activity could be detected in the filter paper fraction. It is known\cite{R7}from Ohtsuki et al.'s high pressure liquid chromatography (HPLC) work that the $^{7}$Be radioactivities observed in the organic phase come from molecules having similar mobility as C$_{60}$ molecules implying $^{7}$Be is somehow attached to fullerene C$_{60}$ molecule. So the possibilities are $^{7}$Be activities in organic phase are coming from either endohedral $^{7}$Be@C$_{60}$ complex or heterofullerene complex or exohedral $^{7}$Be-C$_{60}$ complex. Since only elements of group IVb and Vb of periodic table are known\cite{R9} to form heterofullerene compounds, so beryllium is not expected to form a heterofullerene compound with C$_{60}$. $^{7}$Be ions stopped in the interstitial positions forming exohedral complex with C$_{60}$ fullerene are not expected to form strong covalent bond with carbon atoms of C$_{60}$ and should be readily dissolved in the hydrochloric acid and remain in the aqueous fraction. Therefore, as concluded earlier\cite{R7} also, the $^{7}$Be activities observed in the organic phase should be associated only with the formation of endohedral $^{7}$Be@C$_{60}$ fullerene complex. So the yield of radioactive endofullerene can be accurately determined from the ratio of activities present in the organic and aqueous fractions. We found that the yield of endohedral $^{7}$Be@C$_{60}$ i.e. the probability of insertion of $^{7}$Be in C$_{60}$ by nuclear implantation technique is (5.6$\pm$0.45)$\%$, where the estimated uncertainty includes both the statistical and systematic errors. 

The radiochemically separated organic fraction containing endohedral $^{7}$Be@C$_{60}$ complex was dried in a plastic crucible. Two HPGe detectors having efficiency $\approx$30$\%$ were used to count endohedral $^{7}$Be@C$_{60}$ and $^{7}$Be in Au samples along with a $^{133}$Ba source for 3 months following identical procedure as described earlier. However the amount of extracted endohedral $^{7}$Be@C$_{60}$ complex was only 5.6$\%$ and so the statistical uncertainty on one day count of 478 keV $\gamma$-ray from endohedral complex was rather large (initially about 4.5$\%$ and 14$\%$ at the end of the run) for each detector. We finally obtained from our measurement that the half-life of endohedral $^{7}$Be@C$_{60}$ complex is shorter than that of $^{7}$Be in Au by (3.3$\pm$2.3)$\%$. Recently Ohtsuki et al.\cite{R19} measured half-life of endohedral $^{7}$Be@C$_{60}$ complex with high accuracy and found that the half-life of endohedral $^{7}$Be@C$_{60}$ complex is shorter than that of $^{7}$Be in Au by (1.20$\pm$0.12)$\%$, in agreement with our measured half-life difference(3.3$\pm$2.3)$\%$ within one standard deviation. 

Let us first try to get a qualitative understanding of our results in terms of electron affinity. We have found experimentally that an implanted $^{7}$Be ion has a very low probability (5.6$\pm$0.45)$\%$ of entering the fullerene cage and they mostly stay in interstitial space forming exohedral $^{7}$Be-C$_{60}$ complex.  
The electron affinities of a 
fullerene molecule(C$_{60}$) and gold atom are
2.6 eV \cite{R20} and 2.3 eV \cite{R21} respectively. 
Both gold and fullerene have
face-centered cubic (FCC) lattice structure, but the lattice parameter of fullerene C$_{60}$ lattice is much larger(14.17 {\AA}) than that(4.08 {\AA}) of Au lattice. So if an implanted $^{7}$Be ion occupies the same geometrical position in both C$_{60}$ and Au lattices, then its distance from the center of the nearest Au atom and C$_{60}$ molecule would be very different. When $^{7}$Be would occupy the octahedral site of Au lattice, then its distance from the nearest Au atom would be $\approx$2{\AA}, whereas in the case of C$_{60}$ lattice, the corresponding distance would be $\approx$7{\AA} and the distance from the nearest carbon atom of C$_{60}$ molecule would be $\approx$3.7{\AA}. Hence $^{7}$Be should retain a significantly higher fraction of its 2s electrons in C$_{60}$ lattice and so the decay rate of $^{7}$Be should be significantly faster in C$_{60}$ lattice than in Au lattice. So the observation of the same(within 0.2$\%$) half-life of $^{7}$Be in C$_{60}$ and Au lattices should imply different geometrical positions for the implanted $^{7}$Be ions in the two lattices. It should also imply that the bond lengths of $^{7}$Be-C$_{60}$ (i.e. the distance between $^{7}$Be and the nearest carbon atom of C$_{60}$)and $^{7}$Be-Au are about equal.

In order to understand the results quantitatively, we have done tight binding linear muffin-tin
orbital(TB-LMTO) method calculations \cite{R2,R5,R6,R22} to determine
the average number of 
2s electrons of $^7$Be for both endohedral and exohedral fullerene complexes. A fullerene molecule consists of 60 carbon atoms placed on a sphere of radius 3.54 {\AA}. These fullerene 
C$_{60}$ molecules are arranged in a face-centered cubic structure
with 14.17 {\AA} lattice constant. For both endohedral and exohedral complex, $^{7}$Be should go to equilibrium positions where the total energy of the system has local minima and the calculated decay rate of $^{7}$Be at those positions should agree with the experimental results.

However we found that TB-LMTO code\cite{R22} was not very suitable for the determination of equilibrium positions in the fullerene C$_{60}$ lattice, because its spheridization of potential requires filling up of the available empty space with close packed empty spheres and since such division is not unique, it introduces an uncertainty in the determination of the total energy of the system. In the case of fullerene C$_{60}$ lattice, this uncertainty could be of the order of 0.4$\%$ in some cases and this is unacceptably large because the differences in binding energies of $^{7}$Be at different sites are very small (of the order of 10 eV). The uncertainty in the determination of the charge density and number of 2s electrons of $^{7}$Be could be up to ten percent in some cases, but this is acceptable for the purpose of estimating the change in half-life of $^{7}$Be and comparing with the experimental results. So the equilibrium position of $^{7}$Be in endohedral $^{7}$Be@C$_{60}$ was taken from a previous density functional calculation\cite{R13} and TB-LMTO code was used to determine the average number of 2s electrons of $^{7}$Be at that position.
  
Lu et al.\cite{R13} performed density functional calculations for endohedral Be@C$_{60}$ and found that the equilibrium position of beryllium ion should be exactly at the center of C$_{60}$ cage. So we performed TB-LMTO calculation placing a $^{7}$Be atom atom exactly at the center of fullerene C$_{60}$ cage. The calculation was performed assuming such endofullerenes will form a face-centered cubic lattice with lattice parameter 14.17 {\AA}. Empty spheres were placed in the interstitial space and one inside the cage for close packing. TB-LMTO calculation shows that the radius of the muffin-tin sphere of the $^{7}$Be atom at the center of C$_{60}$ cage is $\approx$3 {\AA} and it almost fills up the entire space inside the C$_{60}$ cage. Let $\Psi _{total}$ be the complete electronic wave function and 
$ \Psi _{Be2s}$ be beryllium 2s state wave function. Then the square of 
the overlap of $ \Psi _{total}$ with $ \Psi _{Be2s}$ i.e
$|< \Psi _{total}|\Psi _{Be2s}>| ^{2}$ represents the average number of 
2s electrons in beryllium ion. Our TB-LMTO calculation shows that the average number of 2s valence electrons of $^{7}$Be atom placed at the center of a C$_{60}$ cage is = 1.07. 

We also calculated the average number of 2s electrons of $^{7}$Be in Au. As shown earlier\cite{R6},in the case of implantation of $^{7}$Be in Au lattice, $^{7}$Be should go to octahedral and tetrahedral sites of Au lattice. TB-LMTO calculations were performed placing a $^{7}$Be atom at octahedral and tetrahedral positions of a face- centered cubic Au lattice having lattice parameter 4.08 {\AA}. The average number of  2s electrons $|<\Psi _{total}|\Psi _{Be2s}>| ^{2}$ 
was found to  be 0.54 and 0.36 for $^7$Be occupying 
octahedral and tetrahedral positions of gold lattice respectively. We expect that as a
result of random implantations, the 
number of $^7$Be atoms occupying tetrahedral sites
would be twice that of occupying octahedral sites, because 
the number of tetrahedral sites are twice that of octahedral 
sites in a face-centered cubic lattice. Hence taking weighted average,
we found $^7$Be
would have 0.42 electrons in its 2s orbital when implanted 
in Au. So the decay rate of $^{7}$Be in endohedral $^{7}$Be@C$_{60}$ should be faster than that of $^{7}$Be in Au. According to Hartree and Hartree's calculation\cite{R23}, the overlap of valence 2s electrons to the total electronic overlap at beryllium nucleus  is only 3.2$\%$ for a neutral beryllium atom. As shown in ref[6], there is a linear relationship between the decay rate of $^{7}$Be and its number of valence 2s electrons. Using Hartree and Hartree's result\cite{R23} along with the linear relationship found earlier\cite{R6}, we obtain that the decay rate of endohedral $^{7}$Be@C$_{60}$ should be faster by $\approx$ 1.1$\%$ compared to that of $^{7}$Be in Au lattice. This result agrees well with our measurement and Ohtsuki et al.'s earlier measurement\cite{R19} that the decay rate was faster by (1.20$\pm$0.12)$\%$. 

We have found that the measured probability of formation of endohedral $^{7}$Be@C$_{60}$ complex by nuclear implantation technique is only 5.6$\%$ and the remaining 94.4$\%$ of the time, $^{7}$Be goes to the interstitial space forming exohedral $^{7}$Be-C$_{60}$ complex. In order to calculate the decay rate of $^{7}$Be when it goes to interstitial space of C$_{60}$ lattice, we have to first determine its equilibrium positions in the fullerene C$_{60}$ lattice. 

Although we have not come across any calculation for determining equilibrium positions of $^{7}$Be in a fullerene C$_{60}$ lattice, there are a large number of calculations\cite{R10, R11, R12, R14} for equilibrium geometries of isolated exohedral fullerene C$_{60}$ complex. From those calculations, it is known that the typical bond length between a carbon and the other ion forming exohedral fullerene C$_{60}$ complex is $\approx$ 2 {\AA}. The distance of the other ion from the center of the C$_{60}$ cage is usually\cite{R10, R11, R12, R14} around 5.5 {\AA} and the equilibrium position of the other ion is most likely to be along a line bisecting normally a C-C bond of C$_{60}$ molecule. Using this information, we carried out TB-LMTO calculations for beryllium in C$_{60}$ lattice and found that there were total energy minima postions when beryllium was on any face of the face-centered cubic C$_{60}$ lattice between a corner C$_{60}$ molecule and a face-centered C$_{60}$ molecule along a line bisecting normally C-C bond of a C$_{60}$ molecule. At equilibrium position, the distance of beryllium ion from the center of C$_{60}$ molecule is about 5.3 {\AA} and C-Be bond length comes out around 2 {\AA}. We find from TB-LMTO calculation that at this position, the average number of valence 2s electrons of beryllium ion is = 0.44. So using Hartree and Hartree's result\cite{R23} along with the experimental observation that only 5.6$\%$ of the implanted $^{7}$Be ions form endohedral complex, the decay rate of the implanted $^{7}$Be in fullerene C$_{60}$ should be faster by $\approx$ 0.1$\%$ compared to that of $^{7}$Be in Au. This result agrees with our experimental observation that the decay rates of $^{7}$Be in fullerene C$_{60}$ and Au are equal within 0.2$\%$.

If we assume that $^{7}$Be ions go to the octahedral or tetrahedral sites of face-centered cubic C$_{60}$ lattice, then according to TB-LMTO calculations, the average number of 2s electrons of $^{7}$Be should be 1.12 and 1.51 respectively. So using the method of ref[6], the decay rate of $^{7}$Be should be faster than that of $^{7}$Be in Au by 1.16$\%$ and 1.8$\%$ respectively, in complete disagreement with our experimental observation. Moreover if $^{7}$Be is at octahedral site, then its distance from the center of the nearest C$_{60}$ molecule would be about 7 {\AA} and so the length of shortest C-Be bond would be about 3.7 {\AA}. Similarly when $^{7}$Be is at tetrahedral site of C$_{60}$ lattice, then its distance from the center of the nearest C$_{60}$ molecule is 6.1 {\AA} and the length of shortest C-Be bond is about 3 {\AA}. At this position, the line joining the center of the C$_{60}$ molecule and $^{7}$Be ion does not bisect C-C bond of C$_{60}$ molecule, but passes through the center of the hexagonal opening of C$_{60}$ molecule. It is also expected\cite{R14} that for most stable exohedral configuration, a beryllium atom should go to a site of high electron density. Our TB-LMTO calculations show that the electron density at tetrahedral site of C$_{60}$ lattice should be about 50$\%$ more than that at octahedral site. However the electron density on a face of the lattice midway between two C$_{60}$ molecules is about five times the electron density at the octahedral site. So from these qualitative arguments, we expect that the octahedral and tetrahedral sites should not be the equilibrium positions of $^{7}$Be in fullerene C$_{60}$ lattice. However our TB-LMTO calculations for the total energy of the system do not support this expectation, perhaps due to the model limitations mentioned above. 

In the case of endohedral $^{7}$Be@C$_{60}$ complex formation, although  C-Be bond length would be 3.54 {\AA} since $^{7}$Be is expected to be at the center of fullerene C$_{60}$ cage, however at that position, $^{7}$Be is equidistant from all 60 carbon ions and is held there by sixty C-Be bonds. So even if each bond is weak, there are sixty such bonds holding the beryllium ion.     

In summary, we have found experimentally that the decay rate of $^{7}$Be in fullerene C$_{60}$ is equal to that of $^{7}$Be in Au within 0.2$\%$. A radiochemical analysis of the irradiated sample shows that only 5.6$\%$ of the implanted $^{7}$Be ions form endohedral $^{7}$Be@C$_{60}$ complexes and the remaining 94.4$\%$ of the $^{7}$Be ions form exohedral $^{7}$Be-C$_{60}$ complexes. The measured decay rate of endohedral $^{7}$Be@C$_{60}$ complex is in agreement with recent accurate measurement of Ohtsuki et al.\cite{R19}. Our TB-LMTO analysis shows that for endohedral $^{7}$Be@C$_{60}$ complex, $^{7}$Be ion should go to the center of C$_{60}$ cage. For exohedral $^{7}$Be-C$_{60}$ complex, we expect that $^{7}$Be should go to one of the faces of face-centered cubic C$_{60}$ lattice forming exohedral complex with C-Be bond length $\approx$2 {\AA}.

We acknowledge useful discussion with R. Vandenbosch (University of Washington, USA)
,Ove Jepsen (Max Planck Institute, Germany) and J. Lu (Peking University,
Beijing, China), G. P. Das (IACS, Kolkata, India) and L. M. Ramaniah (Bhabha Atomic Research Centre, Mumbai, India) and A. Mookerjee (S. N. Bose Centre for basic sciences, Kolkata, India).
\begin{figure}
\caption{$\gamma$-ray spectrum from the decay of $^7$Be implanted in C$_{60}$ 
fullerene}
\label{f1}
\end{figure}

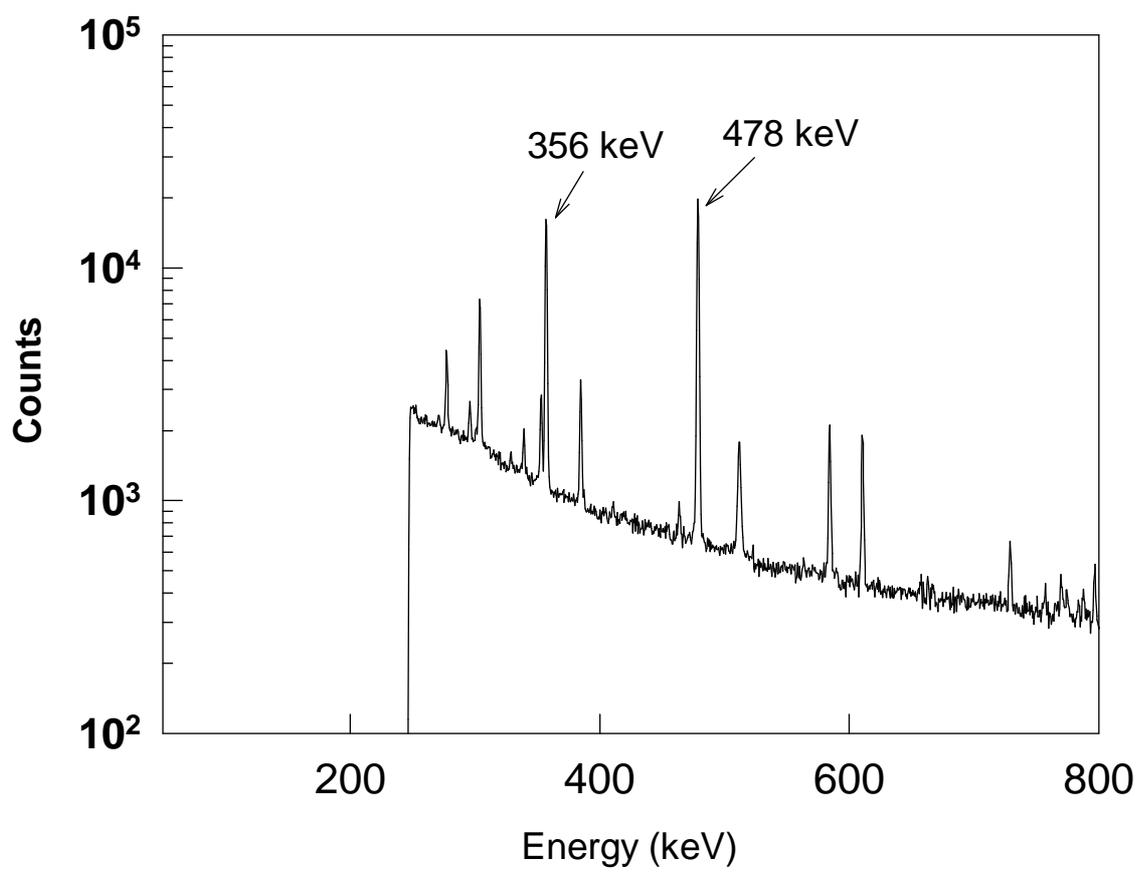
\end{document}